\documentclass[preprint,showpacs,preprintnumbers,amsmath,amssymb]{revtex4}
\usepackage{graphicx}% Include figure files
\usepackage{dcolumn}% Align table columns on decimal point
\usepackage{bm}% bold math
\newcommand{\bfi}[1]{\mbox{\boldmath $#1$}}
\newcommand{\bfis}[1]{\mbox{\boldmath ${\scriptstyle #1}$}}
\newcommand{\vb}{{\bfi b}}

\newcommand{\vs}{{\bfi s}}
\newcommand{\vib}{{\bfis b}}

\newcommand{\viq}{{\bfis q}}

\begin{document}

\title{Monte Carlo integration in Glauber model analysis of 
reactions of  halo nuclei}

\author{
K. Varga$^{1}$\footnote{(e-mail: vargak@ornl.gov).},
Steven C. Pieper$^2$\footnote{(e-mail: spieper@anl.gov)},
Y. Suzuki$^3$\footnote{(e-mail: suzuki@nt.sc.niigata-u.ac.jp)}, and R. B. Wiringa$^2$\footnote{(e-mail: wiringa@anl.gov)}}

\affiliation{
$^1$ Solid State Division, Oak Ridge National Laboratory, Oak Ridge,
37831 Tennessee, USA
\\
and
\\ 
Institute for Nuclear Research of the Hungarian Academy 
of Sciences (ATOMKI), 4000 Debrecen, Hungary PO BOX 51
\\
$^2$ Argonne National Laboratory, Argonne, IL 60439, USA
\\
$^3$ Department of Physics, Niigata University, Niigata 950-2181, Japan}

\vspace{1cm}

\begin{abstract}
Reaction and elastic differential cross sections are 
calculated for light nuclei
in the framework of the Glauber theory. The optical phase-shift function
is evaluated by Monte Carlo integration. This enables us to use
the most accurate wave functions and calculate the phase-shift functions
without approximation. Examples of proton nucleus (e.g. p-$^6$He, p-$^6$Li) 
and nucleus-nucleus (e.g. $^6$He$-^{12}$C) scatterings illustrate 
the effectiveness of the method. This approach 
gives us a possibility of a more stringent analysis of 
the high-energy reactions of halo nuclei.
\end{abstract}

\maketitle

\section{Introduction}

  The study of hundreds of new unstable nuclei 
has become possible in the 
new radioactive beam facilities. The measurement of the 
cross sections at high energies is an important 
experimental tool for studying these elements \cite{tanihata85}. 
The observed interaction
cross sections can be related to the wave functions of these nuclei 
through the Glauber theory \cite{Glauber}; thus one can 
obtain information on 
the structure of these exotic isotopes. This direct relation 
between the structure and reaction cross section  is, 
however, hampered by 
several problems. The first and most serious one is that the 
calculation of the cross section by using accurate nuclear 
wave functions is difficult. Also the 
solution of the nuclear many-body problem is notoriously 
complicated. 

   In Glauber theory the nucleus-nucleus elastic scattering amplitude 
is obtained by integrating the optical phase-shift function over
the impact parameter. The optical phase-shift function is the matrix 
element of the multiple-scattering operator between the product of
wave functions of the target and projectile. The difficulty of the evaluation 
of this matrix element stems from the fact that 
the  multiple-scattering operator is an $A$-body 
($A$ is the total number of nucleons involved) operator, 
a product of pairwise nucleon-nucleon scattering operators. 
To avoid the calculation of the matrix element of an $A$-body operator 
several simplifying approximations have been introduced over the years.
One popular method is to replace the wave function of the target
or both the projectile and the target by densities.
This is then further simplified by assuming that the density of the nucleus
is a product of one-body densities. In that case the matrix element
of the multiple-scattering operator becomes a product of two-body
matrix elements and is easy to calculate. Another popular approach is to 
use an inert core and concentrate on the effect of the valence particles 
only. One can also expand the multiple-scattering operator as a sum of 
two-, three-, ..., $A$-body operators and truncate the expansion at some
level. The most popular method is the ``optical-limit approximation'' (OLA)
where the complicated multiple-scattering operator is replaced by a simple
two-body operator in a somewhat {\it ad hoc} way. 

  The problem of these approximations is that as the ``exact'' scattering
amplitude is not available, it is very difficult to judge how good 
they are. Some of the assumptions sound very serious; the usage of one-body
densities most definitely washes out the effects of nuclear correlations.
Some of these approximations may be especially bad for weakly bound 
halo nuclei as has been discussed by many 
authors~\cite{bertsch,ogawa,al96,yabana,al97}.

  To avoid the complication of the analytical calculation of the cross
section we will use Monte Carlo integration. 
We will show that the accuracy of this approach is very good and 
one can obtain very reliable
results with very little effort compared to the previous approaches.
Moreover, unlike the previous calculations, no approximation 
has to be introduced in the evaluation of the Glauber amplitude. 
We use many-body wave functions directly without loosing 
delicate correlation effects.  
In the framework of the
Glauber theory, Monte Carlo integration has been used to calculate 
the nuclear transparency in the 
$\gamma n \rightarrow \pi^- p$ process in $^4$He and $^{16}$O
in Ref. \cite{GHP96}.

  This approach also allows us to use the most advanced quantum
Monte Carlo (QMC) wave functions available for light nuclei. These
wave functions are obtained by the solution of the nuclear Hamiltonian
with realistic two- and three-body interactions. These {\it ab initio} QMC
calculations of the energy spectra and other nuclear 
properties are in good agreement with experiments 
in the $A \leq 8$ region \cite{QMCCAL}. The QMC predicts the radii of the light nuclei
quite reliably so that a direct comparison of the nuclear sizes and reaction
cross sections becomes possible.

  The road map of this paper is as follows. First we 
introduce the most important quantities of the Glauber theory in Sec. II. 
Next we show some of the details of the calculation in Sec. III
and of the wave functions in Sec. IV. 
The presentation of the results can be found in Sec. V. 
A brief summary and outlook closes the paper in Sec. VI. 

\section{Glauber theory}

Glauber's multiple-scattering theory \cite{Glauber} of high-energy 
collisions, which is based on the eikonal and adiabatic approximations, is 
widely accepted as a standard tool for calculating various cross sections.
The probability that two colliding nucleons in the projectile and 
the target lead to the excitation of the nuclei is almost unity in 
high-energy reactions (at several hundred MeV/$A$). Since the nuclear 
force is short-ranged, the probability of nuclear excitation 
in the collision reflects the geometrical size of the nuclei. 
The interaction or reaction cross section can thus be related 
to the size and structure of the nuclei. 

The basic quantity in the Glauber theory is 
the optical phase-shift function $\chi_{\rm el}(\vb)$ defined by 
\begin{equation}
{\rm e}^{i\chi_{\rm el}(\vib)}=
\langle \Phi^{\rm P}_0 \Phi^{\rm T}_0 \vert
\prod_{i\in {\rm P}} \prod_{j\in {\rm T}} 
[ 1-\Gamma(\vb+\vs^{\rm P}_i-\vs^{\rm T}_j)]
\vert \Phi^{\rm P}_0 \Phi^{\rm T}_0 \rangle,
\label{NN_Opt}
\end{equation}
where ${\bfi b}$ is the impact parameter, and 
$\Phi^{\rm P}_0$($\Phi^{\rm T}_0$) is the intrinsic 
projectile(target) wave function 
with its center-of-mass wave function removed. 
The profile function $\Gamma$ is a two-dimensional 
Fourier transform of the nucleon--nucleon scattering amplitude
\begin{equation}
f(\theta, \phi) = \frac{ik}{2\pi} \int\!\! d\vb \,
{\rm e}^{-i\viq \cdot \vib} \Gamma(\vb).
\label{nn_SA}
\end{equation}
In Eq.~(\ref{NN_Opt}) one integrates over all (independent)
intrinsic coordinates ${\bfi r}^{\rm P}_i$ and ${\bfi r}^{\rm T}_j$.
The two-dimensional vectors, ${\vs}^{\rm P}_i$ and ${\vs}^{\rm T}_j$,
are the projections of ${\bfi r}^{\rm P}_i$ and ${\bfi r}^{\rm T}_j$ 
onto the $xy$-plane which is perpendicular to the incident direction 
of the projectile.

The nucleus--nucleus elastic scattering amplitude is easily obtained 
once the optical phase-shift function is available:
\begin{equation}
f_{\rm el}(\theta, \phi) =
\frac{iK}{2\pi} \int \! d\vb\, {\rm e}^{-i\viq \cdot \vib}
[1 - {\rm e}^{i\chi_{\rm el}(\vib)}],
\label{NN_SA}
\end{equation}
where $K$ is the wave number of the relative motion between the two nuclei. 
The effects of the Coulomb interaction can be easily incorporated as well,
but, as they are important only at extreme forward angles, 
we omitted the Coulombic contribution in the present analysis. 
The total reaction cross section 
is obtained by subtracting the elastic cross section from the total cross 
section: 
\begin{equation}
\sigma_{{\rm reac}}=\int{d{\bfi b}\,\left(1-\big|{\rm e}^{i\chi_{\rm el}(\vib)}
\big|^{2}\right)}.
\label{reacx}
\end{equation}

The Glauber theory is a non-perturbative theory;
its strength is that it directly employs information on
the bare nucleon--nucleon interaction and thereby makes it 
possible to relate the reaction dynamics to the underlying 
interaction of the constituent particles. 
At lower energies and for heavier nuclei than those considered in this paper,
the bare nucleon-nucleon cross sections must be corrected for
medium effects related to Pauli blocking and effective
masses\cite{PP92}.
Usually $\Gamma$ is 
not calculated from the bare nucleon--nucleon interaction but 
parametrized in a convenient form (like Gaussians) 
so as to fit the empirical nucleon--nucleon scattering amplitude 
through Eq.~(\ref{nn_SA}):

\begin{equation}
\Gamma(\vb)={\sigma_\tau \over 4\pi \beta_\tau}
(1-i\alpha_\tau)\, {\rm exp}\left(-\frac{{\bfi b}^2}{2\beta_{\tau}}\right)
\label{gammaform}
\end{equation}
where the parameters depend on the isospin of the nucleons
($\tau$=pp, nn, pn). Other operators, like spin-orbit and tensor,
may also be necessary (especially if the spins of the target 
and projectile are nonzero), but those are most often neglected. 
Their inclusion would not cause any problem in the present approach.

As seen in Eq.~(\ref{NN_Opt}), the optical phase-shift function is 
defined by a many-body multiple-scattering operator and obviously 
its calculation is very involved. One often uses some simplification 
like the OLA, which is just the first-order approximation
in the cumulant expansion \cite{Glauber}:

\begin{equation}
{\rm e}^{i\chi_{\rm OLA}({\bfis b})}
={\rm exp}\left\{-\int \! \int d{\bfi r}^{\rm P} d{\bfi r}^{\rm T} 
\rho^{\rm P}({\bfi r}^{\rm P})\rho^{\rm T}({\bfi r}^{\rm T})
\Gamma({\vs}^{\rm P}-{\vs}^{\rm T}+\vb)\right\},
\label{gptola}
\end{equation}                    
where e.g., $\rho^{\rm P}({\bfi r}^{\rm P})$ is the single-particle 
density of the projectile nucleus. 

Several authors have shown that a treatment beyond the OLA 
is necessary for a quantitative analysis of the 
reaction cross sections \cite{bertsch,ogawa,al96} as well as the 
elastic scattering cross sections \cite{yabana,al97}. 
This is particularly true for loosely coupled systems 
such as halo nuclei because breakup effects are not properly 
accounted for in the OLA. 
There have been considerable efforts which attempt to calculate 
higher-order or all-order terms of the multiple-scattering operator. 
It is known that the matrix elements of 
higher-order terms should be evaluated by two- and more-particle 
densities \cite{Glauber}. Using the density constructed 
from uncorrelated wave functions \cite{czyz,franco,yin,gogary} is thus 
not sufficient to understand the importance of the correlated motion 
in nuclei even if all-order terms are evaluated. 

Very recently a method of calculating the complete Glauber 
amplitude \cite{b99} 
has been proposed and applied to p+$^6$He elastic scattering using 
a microscopic $\alpha$+n+n three-cluster wave function for $^6$He~\cite{arai}. 
It uses an expansion of the multiple-scattering operator  
and evaluates each term analytically provided that the 
profile function $\Gamma$ is expressed by Gaussians.  
The method has been extended to a nucleus--nucleus case as well with 
considerable success \cite{Badawy}. However, this method 
is limited by two things: one is that the number 
of terms in the expansion becomes prohibitively large for heavier 
nuclei; another is that if $\Gamma$ is not a Gaussian then the analytic 
integration for correlated wave functions is hopeless.

\section{Monte Carlo Integration}
Our purpose is to present a powerful method of 
calculating the matrix element of the multiple-scattering operator
completely by taking another route, a Monte Carlo integration (MCI) 
of the multiple-scattering operator. No method has so far been 
available to calculate the Glauber amplitude completely for general 
correlated wave functions. The Monte Carlo integration  seems to be 
the most natural way to calculate the phase-shift function. Its advantages 
are  quite clear: (1) {\sl there is no restriction on 
the form of the target or projectile wave functions 
(general few- or many-body wave function can 
be used)}; (2) {\sl the full multiple-scattering operator can be used 
(there is no need for expansion or truncation)};
(3) {\sl it is very simple compared to earlier calculations}.

In our present approach 
the multiple-scattering operator is assumed to be 
independent of spins, depending only on 
the spatial coordinates and the isospins. The isospin 
dependence arises when different profile functions are used 
between the pp (nn) and pn pairs. 
The wave functions  
$\Phi_0^{\rm P}$ and $\Phi_0^{\rm T}$ are represented by a multicomponent 
(approximately 
%%% $2^A\times \left(\begin{array}{c} A \\ Z \end{array}\right)
$2^A\times (^A_Z)$)
vector in the spin-isospin space. The matrix elements
are calculated by taking the scalar product of the bra and the ket in the 
spin-isospin space and using MCI in the coordinate space.

In MCI, the integration points are generated by the Metropolis
algorithm using $\vert\Phi_0^{\rm P}\Phi_0^{\rm T}\vert^2$ 
as a guiding function. 
One may possibly 
get better accuracy and convergence by using 
$(a+br^2)\vert\Phi_0^{\rm P}\Phi_0^{\rm T}\vert^2$
($r$ is the root mean square radius of the projectile) or 
some similar expression which increases the weight of the asymptotic
part, but as will be shown later the simple 
$\vert\Phi_0^{\rm P}\Phi_0^{\rm T}\vert^2$
form is sufficient for the present purposes.

In the MCI we first generate a set of $N$ integration points 
by a Metropolis random walk and store them. Then the optical phase-shift 
function is 
calculated over these configurations for each discretized value of 
the impact parameter $b$. In this way we not only save computational 
time but we have a ``correlated sampling'' for different impact parameters
avoiding the independent statistical errors of multiple Metropolis 
walks. The reaction cross section or the elastic differential cross 
section is calculated by numerical integration over the impact parameter. 

The impact parameter ${\bfi b}$ 
is a two-dimensional vector which can be parametrized as
($b {\rm sin}\phi,\, b {\rm cos} \phi)$. For spherical nuclei the 
phase-shift function has no dependence on
$\phi$, so the integration in Eqs. (\ref{NN_SA}) and (\ref{reacx}) 
over ${\bfi b}$ 
can be reduced to that over the radial variable $b$.
In the nonspherical case, one has a two-dimensional integration over the impact 
parameter, which increases the number of discretization points. The 
practical example studied in this paper shows that the dependence on $\phi$
is almost negligible. 
 
To test the MCI evaluation of the optical phase-shift function, a 
simple but nontrivial example, $\alpha+\alpha$ 
scattering at 5.07 GeV/$c$  ($T_{\alpha}$=2.57 GeV), 
has been considered. Taking a single harmonic-oscillator
shell-model wave function for the $\alpha$-particle, the phase-shift 
function and the reaction cross section can be analytically 
calculated \cite{b99}. The MCI results are tested against this analytical
example. Figure 1 shows the elastic differential cross section
as a function of four-momentum transfer squared $-t=\hbar^2{\bfi q}^2$ 
using 100,000
MCI points. For small momentum transfer (this is the region where 
the present day experiments can be performed) a very small number of MCI
points is sufficient to get reliable results. For larger momenta 
the oscillatory behavior coming from ${\rm e}^{-i\viq \cdot \vib}$ 
becomes more rapid, requiring more accurate integration for relatively 
large $b$, but even for the $10^5$ points used the computational time
is still almost negligibly small. Table I shows the reaction 
cross section and the mean-square (rms) radius of the $\alpha$-particle 
obtained from this calculation.  Naturally, the analytically 
calculated values are perfectly reproduced provided one uses a sufficiently 
large number of MCI configurations.  These tests show that one can 
confidently use the MCI in calculation of the optical phase-shift
function and related quantities.

\section{Wave functions}

The wave functions, except for $^{12}$C, are 
obtained by the QMC method \cite{QMCCAL,wiringa} using the Argonne 
$v_{18}$ two-body \cite{argonne} and the Illinois three-body interaction IL2
\cite{illinois}. The QMC method with these interactions provides a reliable 
description of the energy levels and different properties of light 
nuclei. Previous calculations are based on cluster, 
few-body, shell-model or mean-field wave functions to calculate the
optical phase-shift functions. These calculations have used schematic
effective interactions with adjustable parameters and other simplifying
approximations. The use of QMC provides us with a realistic, {\it ab initio}, 
microscopic wave function. The $^{12}$C nucleus is not yet accessible in QMC
and for the $^{12}$C a three-$\alpha$  microscopic cluster-model wave 
function is used. In this model the internal wave functions
of the $\alpha$-particles are single shell-model Slater determinants and the
relative motion between the clusters is represented by linear combinations of Gaussians.
The combination coefficients are determined variationally by solving
the 12-nucleon Schr\"odinger equation with an effective 
(Minnesota~\cite{minnesota}) two-nucleon interaction. 

The QMC methods include variational Monte Carlo (VMC) and
Green's function Monte Carlo (GFMC) methods.
The VMC is an approximate method that is used 
as a starting point for the more accurate GFMC calculations.
The VMC method starts with the construction of a 
variational trial function of specified angular momentum, parity and isospin,
$\Psi_T(J^{\pi};T)$, using products of two- and three-body correlation 
operators acting on a fully antisymmetrized set of 
one-body basis states.
Metropolis Monte Carlo integration is used to evaluate 
$\langle \Psi_T | H | \Psi_T \rangle$, giving an upper bound to the 
energy of the state. The GFMC method is a stochastic method that 
systematically improves on $\Psi_T$ by projecting out excited
state contamination using the Euclidean propagation
$\Psi(\tau) = \exp [ - ( H - E_0) \tau ] \Psi_T$.
In the limit $\tau \rightarrow \infty$, this leads to the exact 
ground state energy. Details of the structure calculations can be found in 
\cite{QMCCAL}.

The calculation of expectation values directly using
\begin{equation}
\langle O (\tau)\rangle =
\frac{\langle\Psi(\tau)| O |\Psi(\tau)\rangle}
{\langle\Psi(\tau)|\Psi(\tau)\rangle} \ ,
\end{equation}
is complicated and computationally too demanding. In our calculations we 
have used the approximate expression:
\begin{equation}
\langle O (\tau)\rangle
\approx \langle O (\tau)\rangle_{\rm Mixed}
     + [\langle O (\tau)\rangle_{\rm Mixed} - \langle O \rangle_T]\ ,
\end{equation}
where the ``mixed'' expectation value
between $\Psi_T$ and $\Psi(\tau)$ is:
\begin{equation}
\langle O \rangle_{\rm Mixed} =  \frac{\langle \Psi_T | O |
\Psi(\tau)\rangle}{\langle \Psi_T | \Psi(\tau)\rangle} ,
\end{equation}
and $\langle O \rangle_T$ is just the variational expectation value.
This approximation is very good, provided that the difference between
the VMC trial function $\Psi_T$ and the exact wave function is reasonably small.

\section{Results}
In this section we present our results for total reaction 
and elastic differential cross sections of different 
projectile-target systems. 
The proton, neutron and matter rms radii of the nuclei 
predicted by the structure calculations are shown in Table II.  
The VMC gives a very good wave function for the alpha particle, 
but underbinds the A=6 nuclei. The GFMC energies of $^6$He and $^6$Li
are in very good agreement with experiment. The GFMC improvement 
of the wave function is especially important for the loosely bound
$^6$He.

As a first application we have calculated the differential cross section for 
p+$\alpha$ elastic scattering at $T_{\rm p}$=0.7 GeV (see Fig. 2). 
(The parameters of the profile function (\ref{gammaform}) 
are taken from~\cite{b99}). The agreement 
with the observed cross section~\cite{grebenjuk,alkhazov} 
is perfect, but unfortunately the 
experimental data are only available up to $0.05$ (GeV/$c)^2$. The differential
cross section obtained by a simple harmonic-oscillator $\alpha$-particle 
wave function is also included for comparison. The realistic and simple
shell-model prediction is nearly identical in the experimentally accessible
region. For larger momentum transfer the two wave functions predict 
significantly different cross sections despite the fact that they
give the same nuclear radius. 

Next we consider the $\alpha$+$\alpha$ elastic scattering 
at 5.07 GeV/$c$. 
In this case the experimental data~\cite{berger} are available 
for a wider range of 
momentum transfer (see Fig. 3). While the simple shell-model 
wave function fails to explain the observed data points, the realistic
wave function leads to good agreement with the experiments. 
The disagreement between the simple shell-model and realistic wave function
prediction is almost an order of magnitude. To explain the data with a simple
shell-model wave function one has to increase the harmonic-oscillator
size 
parameter drastically and that would lead to an unrealistic $\alpha$-particle 
radius (about 20\% too large).
This example clearly shows the importance of the realistic wave functions
and the sensitivity of the experimental results to the details of
the nuclear structure.

Table III compares the reaction cross sections calculated for various 
systems with measured interaction cross sections, $\sigma_{\rm int}$. 
The interaction cross section is defined as the sum of the
cross sections for all channels in which the nucleonic composition
of the projectile changes. In high-energy collisions, 
the projectile can only lose nucleons, that is, the probability of   
pickup processes is negligible, and the difference between 
$\sigma_{\rm reac}$ and $\sigma_{\rm int}$ is expected to be small. 
Possible differences between them come from two processes: one 
is the inelastic excitation of the projectile, which may occur 
if the projectile has a particle-bound excited state. Another is a 
process in which the projectile remains in its ground state while the target 
gets excited. These processes contribute to $\sigma_{\rm reac}$ but not 
to $\sigma_{\rm int}$. When the projectile is a halo nucleus that 
has no particle-bound state, like $^6$He, the second process can be ignored 
because the halo nucleus is easily broken by a small 
shock so it is unlikely that it remains in its ground state while 
exciting the target. 
The example of $^6$He+$^{12}$C collision at 0.8 GeV/$A$ bears out 
this argument; the calculated reaction cross section is indeed 
close to the measured interaction cross section. 
The $^6$Li+$^{12}$C reaction cross section is slightly larger than the 
interaction cross section. It remains an open question  
whether this difference can be accounted for by the above processes. 
A similar comment may be applied to the $^{12}$C+$^{12}$C case. 
For the sake of reference the reaction cross sections measured at 
0.87 GeV/$A$ are included in the table. 

Figures 4.a and 4.b present the elastic differential cross sections for 
p+$^6$He and p+$^6$Li at $T_{\rm p}$=0.7 GeV. 
The experimental data \cite{egelhof,alkhazov} are available only up 
to $-t=0.05$ (GeV/$c$)$^2$. In that region (Fig. 5.b) the best
theoretical (GFMC) cross sections slightly underestimate the experimental
data, especially for $^6$He. An $\alpha$+n+n cluster-model 
result~\cite{b99} for p+$^6$He at $T_{\rm p}$=0.717 GeV, which is obtained
by using the wave function~\cite{arai} solved in a restricted model space with 
the Minnesota effective interaction, is also included for comparison. 
The cluster-model result agrees perfectly with the experimental data.  
(See Fig. 2 of Ref.~\cite{b99}.) The matter rms radius 2.51 fm 
in the cluster model, 2.56 fm in the VMC and 2.61 in the GFMC 
calculation. The slight underestimation of the differential cross section
by the GFMC might be due to the fact that the size of $^6$He given by GFMC is 
a little too large. 
For higher momentum transfer, there is a substantial difference between 
$^6$He and $^6$Li as well as the cluster and QMC results.

The full and OLA calculations are compared for the reaction and elastic 
differential cross sections. Table III compares the reaction cross 
section for p+$^6$He and p+$^6$Li at $T_{\rm p}$=0.7 GeV. The OLA 
cross section is slightly smaller than that of the full calculation, 
which, differently from a nucleus-nucleus case, holds true 
for a proton-nucleus system~\cite{ogawa,b99}. Figures 5.a and 5.b compare 
the elastic differential cross sections of p+$^6$He and p+$^6$Li at 
$T_{\rm p}$=0.7 GeV. The difference in the cross sections at small 
four-momentum transfer between 
the full and OLA calculations is magnified in 
Fig. 5.b. Some difference can be seen in the cross section versus $-t$ 
slope for the case of $^6$He or in the magnitude of the cross section 
for the case of $^6$Li at small $|t|$, but the more conspicuous differences appear 
near and beyond the first dip of the cross section.

Differential cross sections versus four-momentum transfer squared 
are plotted in Fig. 6 for elastic scattering of several nuclei 
on a $^{12}$C target at 0.8 GeV/$A$ incident energy. The angular 
distributions show Fraunhofer-type diffraction patterns and the 
first dips move 
to smaller angles with increasing mass number. The basic feature of these 
trends can be understood 
in a strong absorption model (SAM), which is quite reasonable for high-energy 
collisions of nuclei. In an extreme version of the SAM,  
the phase-shift function is assumed to satisfy the relation 
\begin{equation}
{\rm e}^{i\chi_{\rm el}(\vib)}=\Theta(b-R)\equiv
\left\{
\begin{array}{rl}
  0  &  b < R \\
  1  &  b \ge R
\end{array},
\right.
\label{strongabsorption}
\end{equation}
where $R$ is a cut-off radius, and is on the order of 
the sum of the radii of the two colliding nuclei. 
(In reality ${\rm e}^{i\chi_{\rm el}(\vib)}$ has a smooth cut-off 
distribution.) In the 
SAM the incoming flux corresponding to the collision 
with $b<R$ is completely absorbed, while the collision 
with $b>R$ receives no effect. (Note that the Coulomb interaction is 
ignored.) Substitution of 
Eq.~(\ref{strongabsorption}) into Eq.~(\ref{NN_SA}) 
leads to the scattering amplitude of 
$f_{\rm el}(\theta, \phi)\propto J_1(qR)/qR$. The differential cross 
section $|f_{\rm el}(\theta, \phi)|^2$ thus vanishes at the 
zeros of the Bessel function $J_1(qR)$, whose  
zeros occur at $qR \approx 3.83, 7.02,...$. With increasing $A$, 
$R$ in general increases, thus the first zero appears at smaller 
values of $-t$. The value of $R$ 
extracted from the dip in the figure corresponds to the empirical 
radius rather well. The $^6$He and $^{12}$C case does not follow 
this rule as the size of $^6$He is larger than that of $^{12}$C.
This is mainly due to the fact that the sharp cut-off assumption is not good 
for $^6$He because of its halo structure. The momentum transfer 
squared corresponding to the second dip is predicted as 
(7.02/3.83)$^2\approx 3.4$ times that of the first dip 
in the sharp cut-off model.

\section{Summary and Outlook}

The Monte Carlo integration was used to facilitate the evaluation 
of the complete Glauber amplitude involving the multiple-scattering 
operator. There was no need to introduce any {\it ad hoc} 
approximation or assumption in this approach. 
The great advantage of this method is that it enables us 
to use accurate, sophisticated wave functions 
of projectile and target nuclei. A number of calculations 
confirmed that it is possible to directly 
relate the wave functions to reaction cross sections 
measured at high energy.

The calculations presented here focused on light nuclei. 
One can carry out similar investigations for heavier elements 
provided that suitable wave functions are available. One 
very often replaces the wave function of the target nucleus with 
a density distribution. This approximation renders calculations for
 heavier targets ($^{27}$Al or $^{208}$Pb) accessible. In that case 
one has to construct a reliable nucleon-target profile function
to replace the nucleon-nucleon profile function, but the rest 
of the calculation is the same as presented in this work. One further
possible approximation is to use an inert core with 
correlated valence nucleons to construct the wave function of the 
projectile. Such a step might be necessary to investigate the cross 
sections of oxygen or sodium isotopes, for example. 

The experimental results mostly cover the low-momentum transfer region
where the elastic differential cross section is not too sensitive to 
the details of the wave functions and simple models do quite well. 
We have shown, however, that there is a strong discrepancy between 
the predictions of  simple model and realistic wave functions. The
high-energy reactions therefore may give information on the details
of the nuclear structure. For example, the $\alpha+\alpha$ scattering 
at 5.07 GeV/c, where the experimental data is available for larger
momentum transfers, can only be described by using realistic wave 
functions for the alpha particle. One hopes  that the new radioactive beam
facilities will provide us experimental data at higher momentum 
transfer and further tests of the wave functions will become
possible.

\acknowledgments
 The work of K.V. is sponsored by the U.S. Department of Energy under 
contract DE-AC05-00OR22725 with the Oak Ridge National Laboratory, 
managed by UT-Battelle, LLC,  and OTKA grant No. T029003 (Hungary). 
This work was started while K.V. stayed at Niigata in the summer of 
2001 as a fellow of Yamada Science Foundation.
The work of S.C.P. and R.B.W. is supported by the U.S. Department of Energy,
Nuclear Physics Division, under contract No.\ W-31-109-ENG-38. 
The work of Y.S. is supported by a Grant-in Aid for Scientific Research 
(No. 14540249).

\vfill
\newpage

\begin{table}

\caption {Comparison of $\alpha +\alpha $ reaction cross section at 5.07 GeV/$c$ and the rms radius of the $\alpha $-particle calculated analytically and by Monte Carlo integration with 100,000 points. The wave function of the $\alpha $-particle is taken as the simple (0s)$^4$ Slater determinant. 
The statistical error of the Monte Carlo integration is given in parenthesis. }
\begin {tabular}{lll}
Method  & $\sigma_{{\rm reac}}$ (mb) & $\langle r^2 \rangle$ (fm$^2$)\\
\hline 
analytic&       242.91               &             2.250              \\
MCI     &       242.8(7)             &             2.251(3)
\end {tabular}
\end {table}

\begin{table}
\caption {Point proton, neutron and matter rms radii and reaction cross 
sections for the collision with a 0.7 GeV proton.
The Monte Carlo statistical errors for $\sigma _{\rm reac}$ are 1 mb in all cases. }
\begin {tabular}{llllll}
System   &wave function&$r_{\rm p}$ (fm) &$r_{\rm n}$ (fm)&$r_{\rm m}$ (fm) & 
                                            $\sigma _{\rm reac}$ (mb)\\
\hline 
$\alpha$ &VMC          & 1.46 & 1.46 & 1.46 & 100   \\
$^6$He   &VMC          & 1.96 & 2.81 & 2.56 & 163   \\
$^6$He   &GFMC         & 1.92 & 2.87 & 2.61 & 172   \\
                                                 
$^6$Li   &VMC          & 2.47 & 2.47 & 2.47 & 165   \\
$^6$Li   &GFMC         & 2.47 & 2.47 & 2.47 & 166   \\
$^{12}$C &cluster      & 2.36 & 2.36 & 2.36 & 254   \\
\end {tabular}
\end {table}

\begin{table}

\caption {Calculated reaction cross sections. Experimental data are the interaction cross sections taken from \protect \cite {riken}. Cross sections 
with $^{\dagger}$ are the reaction cross sections at 0.87 GeV/$A$\protect\cite{jaros}.
The statistical error of the Monte Carlo integration is given in parenthesis.
The reaction cross section marked by $*$ are obtained in the OLA case.}
\begin {tabular}{llll}
System &Energy (GeV/$A$)&$\sigma _{\rm reac}$(mb)&
$\sigma _{\rm int}$(mb)(Exp.)\\
\hline 
p+$\alpha $         &0.7    & 100(1)& \\
p+$^6$He            &0.7    & 172(1)& \\
p+$^6$Li            &0.7    & 166(1)& \\
p+$^6$He            &0.7    &$166^*$& \\
p+$^6$Li            &0.7    &$164^*$& \\
$^6$He+$^{12}$C     &0.7    & 721(2)& \\
$^6$Li+$^{12}$C     &0.7    & 708(2)& \\
$\alpha $+$\alpha $ &0.6425 & 235(1)& \\
p+$^{12}$C          &0.8    & 261(1)& 262$\pm$13.5$^{\dagger}$ \\
$\alpha $+$^{12}$C  &0.8    & 506(1)& 503$\pm 5$, 527$\pm$26$^{\dagger}$ \\
p+$^6$He            &0.8    & 178(1)& \\
p+$^6$Li            &0.8    & 171(1)& \\
$^6$He+$^{12}$C     &0.8    & 733(2)& 722$\pm 6$ \\
$^6$Li+$^{12}$C     &0.8    & 712(2)& 688$\pm 10$ \\
$^{12}$C+$^{12}$C   &0.8    & 865(1)& 856$\pm 9$, 939$\pm$49$^{\dagger}$
\end {tabular}
\end {table}

\begin{figure}
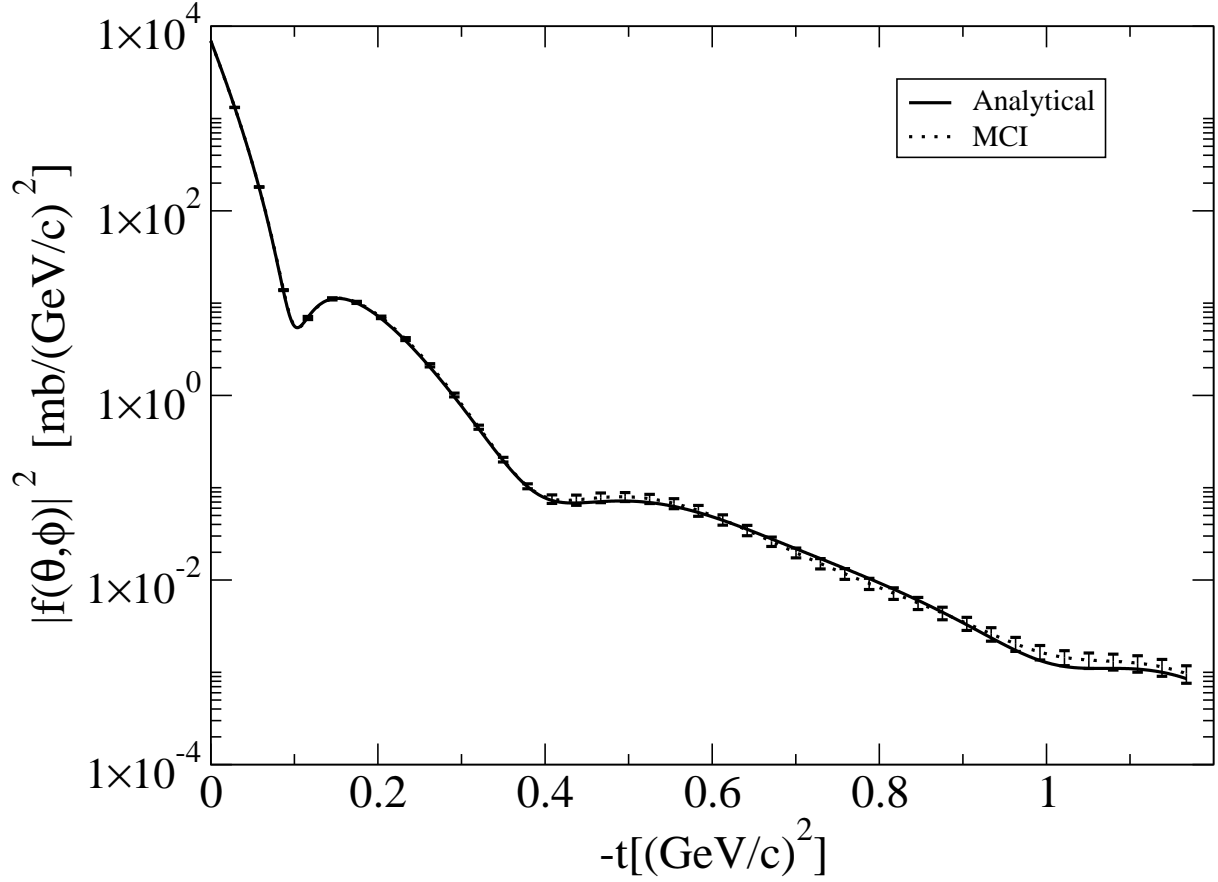


\caption {Center-of-mass differential cross section versus four-momentum transfer squared for $\alpha +\alpha $ elastic scattering at 5.07 GeV/$c$. The solid curve is the result of the analytical calculation, while the result of Monte Carlo integration is shown for 100,000 points.}
\includegraphics *[scale=0.65]{glfig1.eps}
\end {figure}

\begin{figure}
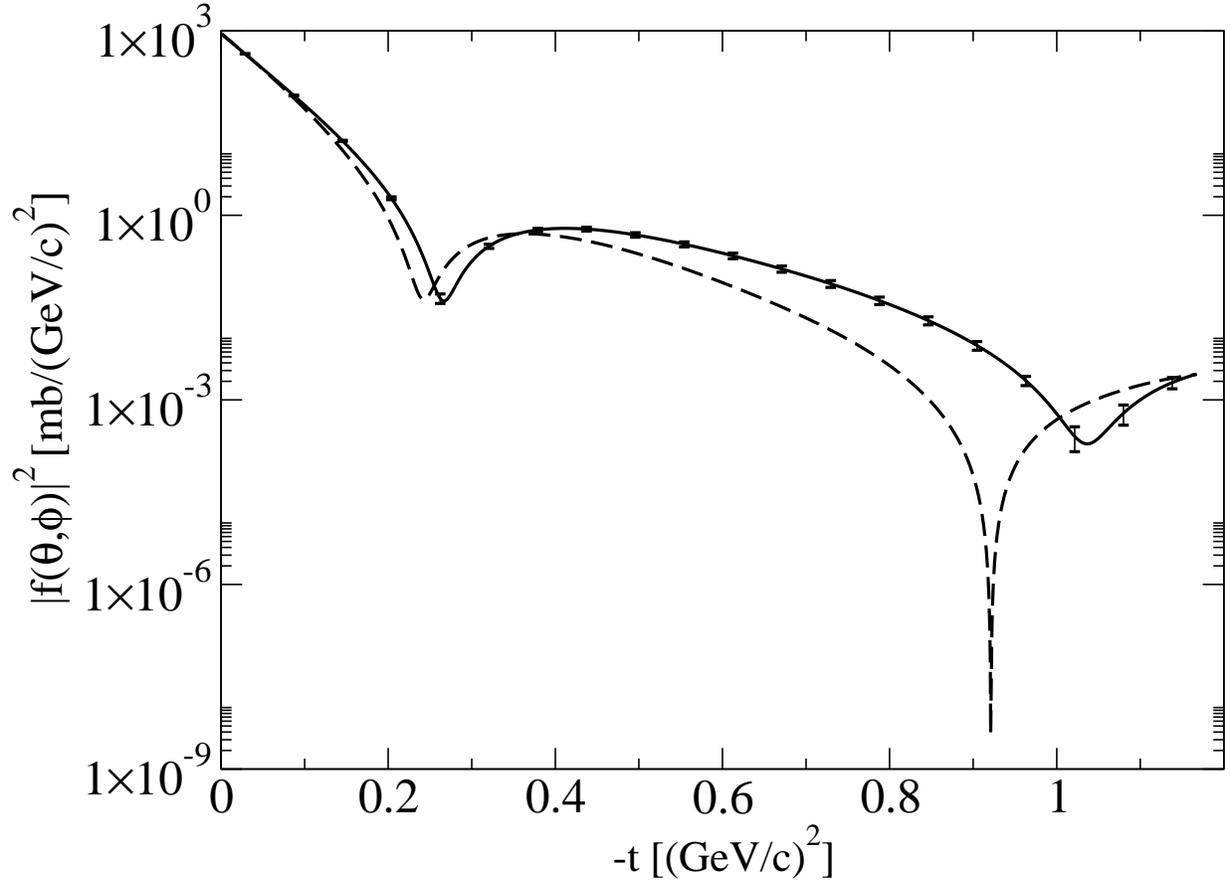


\includegraphics *[scale=0.65]{glfig2.eps}
\caption {Comparison of the differential cross section for p+$\alpha $ elastic scattering at $T_{\rm p}$=0.7 GeV obtained by a simple shell-model (dashed line) and a realistic (solid line) wave function of the $\alpha $-particle. 
The two wave functions give the same rms radius; Monte Carlo 
error estimates are 
shown for the latter case.  The data are available only for $-t \le 0.05$ 
(GeV/$c$)$^2$~\protect \cite {grebenjuk,alkhazov}, but not shown 
because they are hardly 
distinguishable from the theoretical curves.}
\end {figure}

\begin{figure}
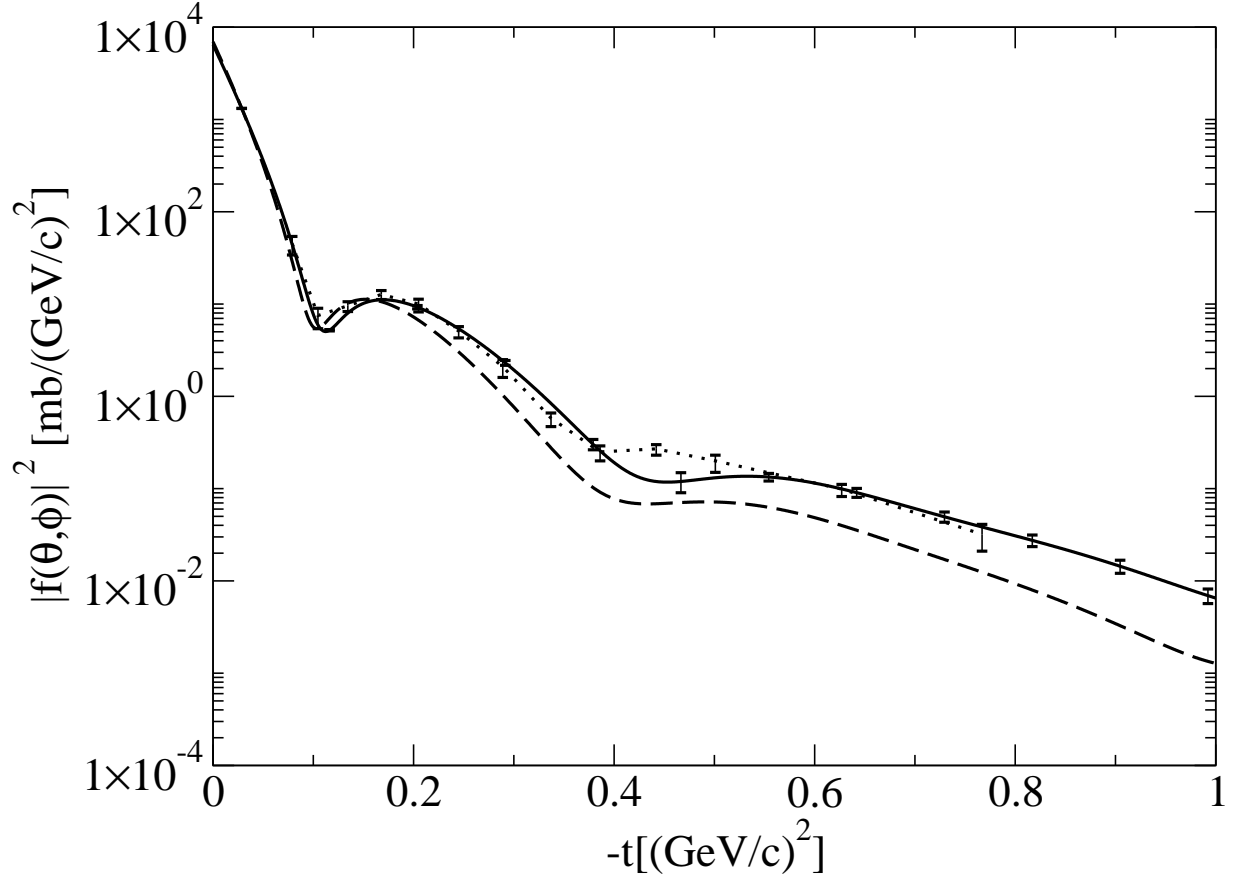


\includegraphics *[scale=0.65]{glfig3.eps}
\caption {Comparison of the experimental and theoretical differential 
cross section for $\alpha +\alpha $ elastic scattering at 5.07 GeV/$c$. The 
solid line shows the results with a realistic $\alpha $-particle wave 
function, 
while the dashed line is obtained by a simple $0s$ shell-model wave function. 
The two wave functions give the same rms radius. The data are taken 
from \protect \cite {berger}. }
\end {figure}

\begin{figure}
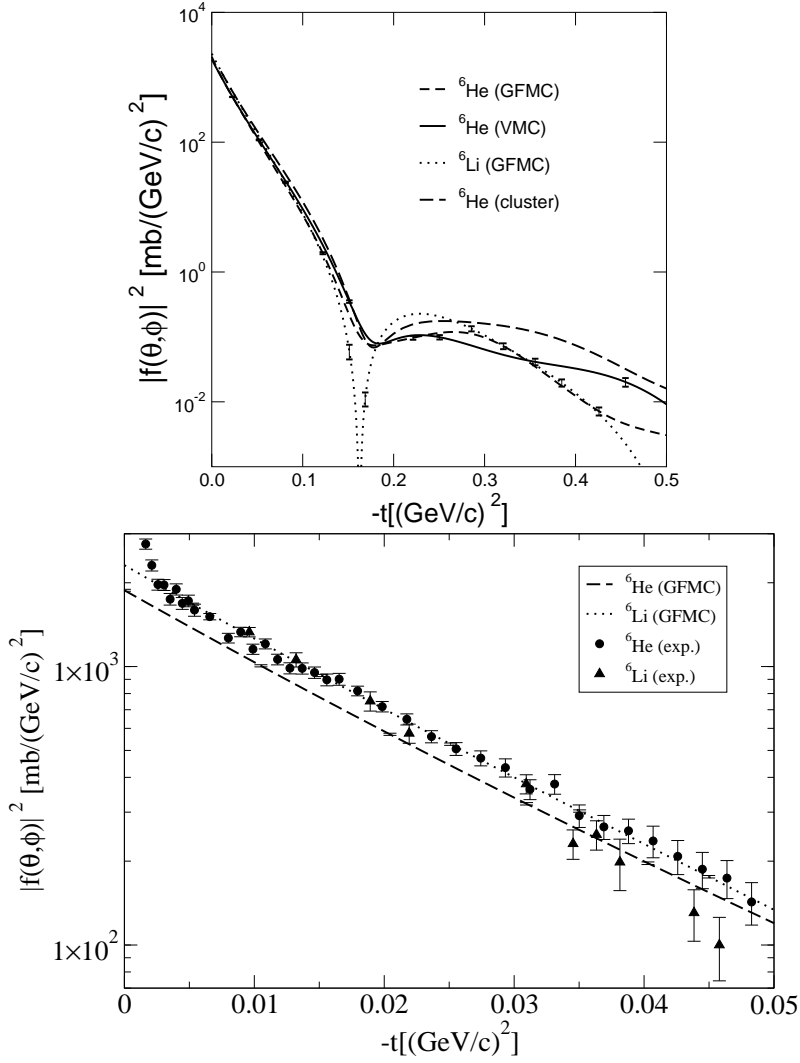


\includegraphics *[scale=0.4]{glfig41.eps}
\includegraphics *[scale=0.4]{glfig42.eps}
\caption {(a) Differential cross section versus 
four-momentum transfer squared for p+$^6$He 
and p+$^6$Li elastic scatterings at $T_{\rm p}$=0.7 GeV. 
An $\alpha$+n+n cluster-model result~\protect\cite{b99} is shown 
for p+$^6$He at $T_{\rm p}$=0.717 GeV. The statistical error 
of the Monte Carlo integration is shown for the GFMC and VMC results. 
(b) Theoretical and experimental \cite{egelhof} differential 
cross sections for  p+$^6$He  and p+$^6$Li elastic scatterings 
at $T_{\rm p}$=0.7 GeV. }
\end {figure}

\begin{figure}
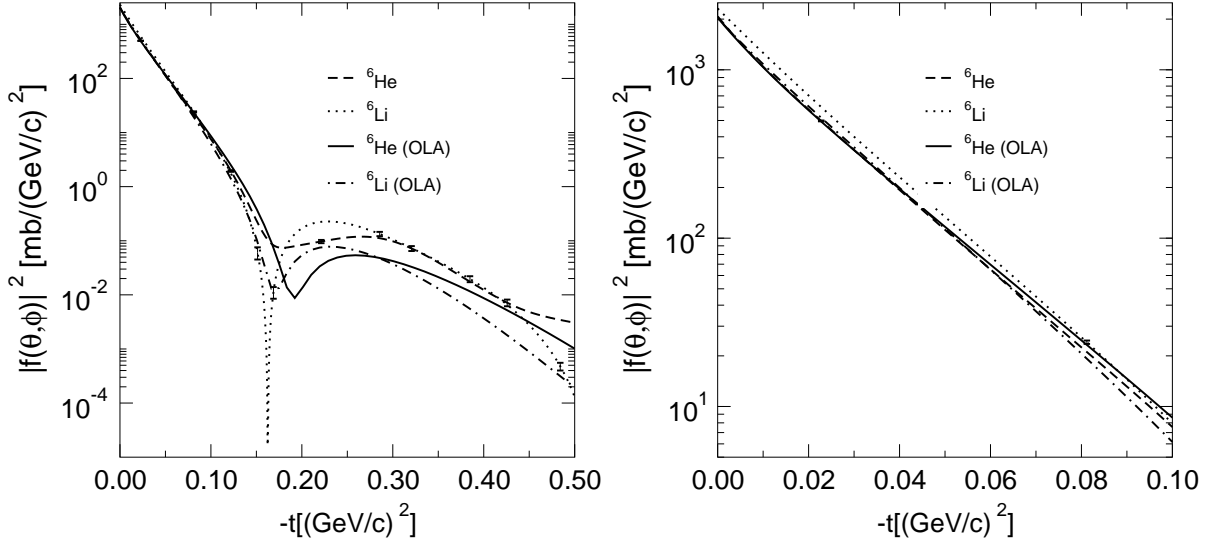

\includegraphics *[scale=0.4]{glfig51.eps}
\includegraphics *[scale=0.4]{glfig52.eps}
\caption {Comparision of the full and OLA differential cross sections. 
The parameters are listed in the caption of Fig. 4. 
The GFMC results are used to calculate the cross sections.} 
\end {figure}

\begin{figure}
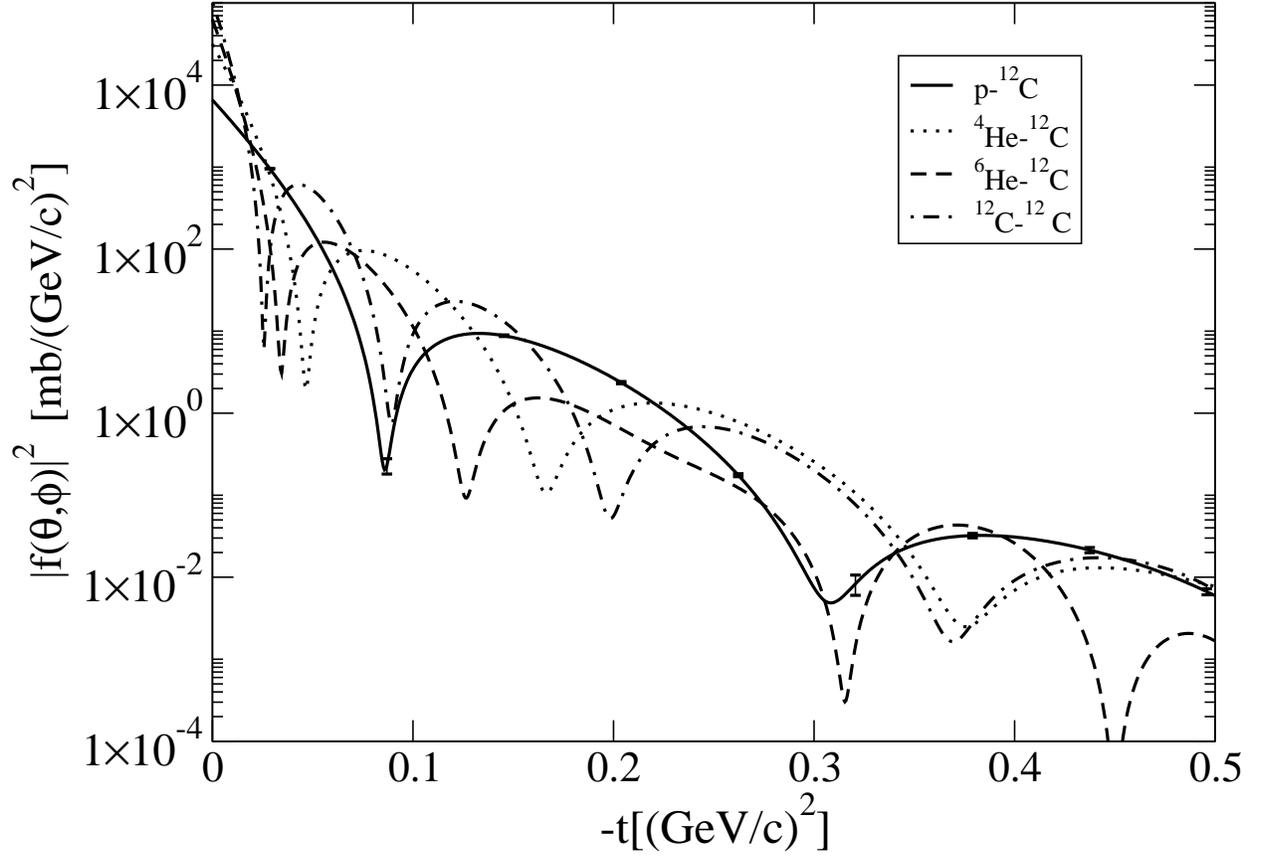


\includegraphics *[scale=0.65]{glfig6.eps}
\caption {Differential cross sections versus four-momentum transfer 
squared for elastic scatterings of p, $\alpha $, $^6$He and $^{12}$C 
on $^{12}$C at 0.8 GeV/$A$.}
\end {figure}

\end{document}